\documentstyle[12pt,epsf]{article}

\begin{document}

\begin{flushright}
SLAC--PUB--8030\\
January  1999  
\end{flushright}

\bigskip\bigskip
\begin{center}
{\bf\large
PROGRAM UNIVERSE and RECENT COSMOLOGICAL RESULTS
\footnote{\baselineskip=12pt
Work supported by Department of Energy contract DE--AC03--76SF00515.}
\footnote{\baselineskip=12pt
Conference Proceedings for ANPA 20 will be available from
ANPA c/o Prof.C.W.Kilmister, \\
Red Tiles Cottage, High Street, Bascombe,
Lewes, BN8 5DH, United Kingdom.}}

\bigskip

H. Pierre Noyes\\
Stanford Linear Accelerator Center\\
Stanford University, Stanford, CA 94309\\
\end{center}
\vfill
           
\begin{abstract}
Recent improvements in astronomical observations 
lead to the conclusion that the Hubble constant lies between 60 and 80
Mpc km$^{-1}$ sec$^{-1}$ and the age of the universe between 11 and 14
Gigayears. Taken together with recent observations of distant type Ia
supernovae and the cosmic background radiation, these limits allow a
check of the consequences of predictions made a decade ago using program universe
and the combinatorial hierarchy that the ratio of baryons to photons
is $1/256^4$ and of dark to baryonic matter is 12.7. We find that the
restrictions on the matter content of the universe and the cosmological
constant are within, and much tighter than, the limits established by
conventional means. The situation is further improved if we invoke
an estimate of the normalized cosmological constant made by E.D. Jones
of $\Omega_{\Lambda} \sim 0.6$. This opens
a ``window of opportunity'' to get the predictions of the ANPA program
in  front of the relevant professional community {\it before} precise
observations lead to a consensus. We urge ANPA members to join us in 
the assault on this breach in the walls of establishment thinking.
\end{abstract}

\begin{center}
Invited paper presented at the $20^{th}$ annual international meeting of the\\
{\bf ALTERNATIVE NATURAL PHILOSOPHY ASSOCIATION}\\
Wesley House, Cambridge, England, September 3--8, 1998\\
\end{center}
\vfill

\newpage

\section{Introduction}

When Fredrick (Parker-Rhodes) discovered the combinatorial 
hierarchy \cite{Parker-Rhodes62}
in 1961, the excitement arose from the successful
calculation of two dimensionless, empirical ratios---the
fine structure constant and the ratio of electromagnetic to
gravitational forces. These numbers were already known to physicists
to better accuracy than his calculation provided, but (then and now)
no extant conventional theory provided a way to calculate them.
These same facts held true for his subsequent calculation of the
electron-proton mass ratio \cite{Parker-Rhodes81a}, and also for many
numbers the ANPA program has produced over the years. I discussed
some of the reasons why established physicists continue to ignore these
results in my introductory lectures presented here a couple of years ago
\cite{Noyes97a}. One basic reason is that the numbers were known
before the calculations were made, leaving the program open to a 
charge of engaging in ``numerology''. 

If I had been lucky, I might have
been able to predict that there are only three generations of neutrinos
before SLAC and LEP demonstrated this experimentally, but I doubt
that this would have made much difference to the reception of our
results by most physicists. The
basic difficulty remains that our line of reasoning is so foreign to
most physicists that any success we have along these lines will 
need to be (a) dramatic, (b) timely, and (c) well publicized in the
relevant professional literature {\it before} the observations are
made. It looks unlikely that these conditions will be met any time
before ANPA 40, so far as particle physics goes.

One reason for this paper is to point out that we may now have a better
chance of getting our cosmological predictions before the relevant
audience in a timely fashion than we have had with particle physics. 
But this ``window
of opportunity'' may easily slip by us unless more effort is put on 
cosmological predictions than we have exerted in the past. Here-to-fore  
the basic observational cosmological
parameters have been so uncertain, and the competing ``conventional'' 
theories so multifarious
and speculative, that our rather precise
results (for those of us who believe in {\it program universe} \cite
{Noyes97b}) seemed to have little prospect of getting attention,
let alone confirmation. This situation has changed quite dramatically
in the last year thanks to a number of different results that 
restrict the Hubble constant to the range 60 to 80 km sec$^{-1}$ 
Mpc$^{-1}$ \cite{Groometal98,Hogan98}, limit the age of the universe
to between eleven and fourteen Gigayears \cite{Groometal98},
require the universe to have much (and perhaps more) of its expansion rate
determined by a repulsive cosmological constant rather than by the 
lack of closure mass, and give direct
measurements \cite{Tyson98} of dark matter between the galactic clusters as well as 
of the dark matter surrounding them. 

The almost totally unexpected result that the universe is
{\it demonstrably} going to keep on expanding forever
was implied by the physical interpretation of {\it program
universe} tabulated in  \cite{Noyes&McGoveran89} a decade ago. 
What I called $M_{vis}$ in that table was meant to imply
``electromagnetically observable'' rather than ``detected by
recording visible light''. This ambiguity  persists in the table
published \cite{Noyes94} in 1994, but the prediction presented there
of the number ratio of baryons to photons, first
published \cite{McGoveran&Noyes91} in 1991, correctly recognizes
that {\it program universe} gives, as a first approximation,
baryonic matter rather than the observational parameter sometimes called
``visible matter''. 

The distinction between baryonic and
observationally visible matter is significant because a possibly substantial
and currently unknown fraction of the baryonic matter may occur
in the form of ``brown dwarfs''. 
What is important here is that our theory gives    
what turns out to be quite a good prediction of the baryon to photon
ratio {\it and} of the ratio of dark to baryonic matter. The energy density 
in photons is known directly from the temperature of the cosmic
background radiation. Further, since our prediction only has two
categories of matter in significant quantities in this epoch, we
get a prediction of the {\it total} matter density independent of
answering the vexed question of how much of the non-visible but
baryonic matter is in the form of brown dwarfs or other ordinary matter.
Thus, knowing the Hubble constant $H_0$, we can predict $\Omega_M\equiv
\rho_m/\rho_c$ where $\rho_c= 3H_0^2/ 8\pi G_N$, $G_N$ being the
Newtonian gravitational constant..   
The absolute mass of the universe we also predicted a decade ago
is still not easily connected to observational data, but even in 1989 pointed
in the direction of an open universe \cite{Noyes&McGoveran89}, in opposition to 
a near consensus among the cosmological theorists. 

As we will show in Section 3, taking seriously our predicted ratio of baryons to
photons and of dark to baryonic matter is already quite
restrictive, and well within the limits allowed by current cosmological
observations.  When augmented by an ``a priori'' estimate of the cosmological
constant made by Ed Jones (cf. Section 4), we end up being able to
make a prediction of the two parameters ($\Omega_M$ and
$\Omega_{\Lambda}$) which specify the
gross cosmology of the universe that is {\it
better} than current observations can test.
The three numbers $M_{Dark}/M_B$, $N_B/N_{\gamma}$ and $M_{\rm Univ}$
play a role in observational
cosmology comparable to the role played by the three Parker-Rhodes 
numbers ($e^2/\hbar c, m_{\rm Planck}/m_{\rm proton}, m_p/m_e$) in particle physics. 
In both observational cosmology and elementary particle physics
the conventional approach requires the numbers to   
be taken from observation and fitted into an hypothesized theory 
rather than calculated from first principles.
In contrast to the empirically well known
Parker-Rhodes numbers,
however, the first two cosmological numbers are only this year beginning to take on 
consensus values at the ten to twenty
per cent level, and $M_{\rm Univ}$ is still know only within a factor of two
or so. This time the ANPA program has a fighting chance
to get our numbers in front of the relevant audience {\it before} they are
well measured rather than {\it after} the fact.

How the ``repulsive cosmological constant'' comes about is another
story, which will only be briefly touched on in this paper. Ed Jones
had reached that result on quite general grounds some time ago, but
unfortunately did not publish his conclusion because of lack of
observational evidence. He is now preparing a short paper on the
subject \cite{Jones98}, which I pray will get into the literature in time to get him
some credit.

But none of my remarks here will make sense unless we have in front of us the
recent observational results which have so dramatically changed the
cosmological picture.

\section{A Brief Survey of Recent Cosmological Results}

A number of factors, which are the culmination of many years of hard
work by many astronomers, astrophysicists, and physical cosmologists,
have converged rather suddenly on definite observational cosmological
results. Partly these are simply the result of the accumulation of data
from the large Keck telescopes in Hawaii and Chile, as well as from
smaller observatories, and from the Hubble Space Telescope. Partly
they are the result of accumulating satellite data, particularly from
Hipparcos and the COsmic Background Experiment (COBE). But the data analysis
would not have been possible without the increasing power and
availability of low cost computers, and would not have yielded such
dramatic results so quickly without some very clever ideas exploiting
the technological and observational opportunities.

Before plunging into my description of some of the results, I wish to
stress that I am an outsider in this field, and have had to rely almost
exclusively on one conference this spring (DM98) and one summer
institute (SSI XXVI) which I had the good fortune to attend.
There was enough discussion of controversial matters in both
of these environs to allow me to believe I could assess reasonably accurately
the outlines of agreement that are emerging, and to draw my own
conclusions. But be warned that my lack of background could have led me
pretty far astray.

One basic fact which seems pretty firm is that Hipparcos has now
supplied a large enough sample of cephid variables with measured
parallax to change the calibration of the distance-luminosity
relation by ten percent or more. So far as the age of the
universe goes, it is equally important that the parallax
of several globular clusters has also been measured.
This turned out to make the oldest objects in our own galaxy 
younger than estimates of the ``age of the universe''
by the right amount to achieve consistency \cite{Perryman98,Groometal98}.
Thus there is no longer an ``age problem''. The universe, and
its contents, have existed something like 12.5 billion 
($1.25 \times 10^{10}$) years, give or take a billion or so;
as our reference time we take the backward extrapolation to
the time when the contents of the universe must have been so
compacted that the question of whether we can trust the laws of physics
enough to extrapolate any earlier becomes, for some of us, the critical
question. Fortunately both ``fireball time''
(when the radiation breaks away from the matter) and the
earlier time of nucleosynthesis (when the neutrons freeze out
and for a few minutes can be used to form deuterons, $^3He$,
alpha particles and $^7Li$) are sufficiently later than this
``epistemological cutoff''  so that we can still
perform relevant laboratory experiments to check our assumptions.
We can remain comfortable with the cosmological
calculations needed in what follows from an operational point of view.
  
This extrapolation back in time starts from ``local'' evaluations
of the ``Hubble Constant''---the distance-velocity relation
between recessional velocity as measured by red-shift using data
out to about 100 Mega-parsec ($1\  Mpc = 3.26...\times 10^6$ light years).
For the nearer galaxies this again depends on the recalibration of the
cephids, but also on getting a handle on the local imhomogenieties
(Virgo cluster, the ``great attractor'', etc., etc.). Recently
much more data has become available on
the ``Peculiar Velocities'' of Galaxies which deviate from the
average Hubble streaming. Consequently one can plot the overall 
distribution of gravitating
matter (rather than the distribution of light) over this enormous---but still ``local''%
---region. These measures have to be self-consistent. When this is achieved,
as is claimed, it reinforces the conclusion, which now comes from
several different types of data, that most of the gravitating matter 
in the universe is dark rather than luminous.

To take the Hubble relation back farther, one needs a ``standard
candle'' that is reliable to as early times as is possible. It turns
out that the type Ia supernovae are numerous enough in the region
where cephid measurements can still be made to collect enough
calibrated light curves to establish what is needed. This took
a very clever combination of physical reasoning and optimal
utilization of resources which have to be shared with many 
other meritorious observational programs. It is this data
which gives firm evidence for a repulsive cosmological constant
\cite{Schwarzschild98}.

Direct measurement of the dark matter itself has been made by detailed
analysis of the defects in the gravitational lenses provided by
clumps of ``local'' galaxies imaging very early galaxies (back to red-shift
5!) \cite{Tyson98}. Most of the lensing comes from the dark matter itself,
not from the sprinkling of visible matter which, presumably, has fallen
into it. Such lenses also exist between visible clumps; these lenses may or
may not include burned out galaxies, but are not optically visible.
These results
confirm the hypothesis that much more of the gravitating matter 
in the universe is dark than luminous. Even with this additional
matter, there is not enough to close the universe in the absence
of an attractive cosmological constant, let alone enough in the
presence of the observed repulsive cosmological constant. This
becomes clear when the COBE data and the type Ia supernovae data
are combined \cite{Glanz98a,Glanz98b}.

\section{Consequences of Two Program Universe 
\hfill\break Predictions}

\subsection{Program Universe}

Here we remind the reader of how we use {\it discrimination}
(``$\oplus$'') between
ordered strings of zeros and ones ({\it bit-strings}) defined by
\begin{equation}
({\bf a}(W)\oplus {\bf b}(W))_w =(a_w-b_w)^2;\ \
a_w,b_w \in 0,1;\ \ w \in 1,2,....,W
\end{equation}   
to generate a growing universe of bit-strings which at each step
contains $P(S)$ strings of length $S$. We use an algorithm known
as {\it program universe} which was developed in collaboration with
M.J.Manthey  \cite{Manthey86,Noyes97b}.
Since no one knows how to construct a ``perfect'' random number 
generator, we cannot in practice start from Manthey's ``flipbit''
(which returns a zero or a one with equal probability when asked), and must
content ourselves with a pseudo-random number generator that, to some
approximation which we will be wise to reconsider from time to time,
will come close to that performance.
Using any available approximation to ``flipbit'' and
assigning an order parameter $i \in 1,2,...,P(S)$ to each string
in our array, Manthey \cite{Manthey86} has given the coding for
constructing a routine ``PICK'' which picks out some arbitrary string 
${\bf P}_i(S)$ with probability $1/P(S)$. Then program universe
amounts to the following simple algorithm:    

\begin{quotation}
PICK any two strings ${\bf P}_i(S)$,${\bf P}_j(S)$, $i,j \in 1,2,...,P$
and compare ${\bf P}_{ij}={\bf P_i \oplus P_j}$ with ${\bf 0}(S)$.

If ${\bf P}_{ij} \neq  {\bf 0}$, adjoin  ${\bf P}_{P+1}:={\bf P}_{ij}$
to the universe, set $P:= P+1$ and recurse to PICK. [This process is
referred to as ADJOIN.]

Else, for each $i \in 1,2,...,P $ pick an arbitrary bit ${\bf a}_i
\in 0,1$, replace ${\bf P}_i(S+1):= {\bf P}_i(S)\Vert {\bf a}_i$,
set $S:= S+1$ and recurse to PICK. [This process is referred to as TICK.]
\end{quotation}

Here the operation ``$\Vert$'' simply extends the string on the left
of the symbol
by adjoining the string to its right (in the instance above, the
arbitrary bit  ${\bf a}_i$ supplied by ``flipbit'')
and adjusting the ordering indices
and resulting string length parameter appropriately. 
We note that any universe
so generated is ``uncrunchable'', to quote John
Wheeler \cite{Wheelerc85}. In our current context this construction,
taken seriously, {\it necessarily} requires that the cosmological
constant be greater than zero, as we will assume below. 

\subsection{Events, Labels, Contents}

This version of {\it program universe}---called ``Program Universe
2'' in the published Ref. \cite{Noyes&McGoveran89}---provides
considerable structure to ``events'', modeled by the two alternatives
presented above. Note that so long as the string
produced by the event is non-null 
(and hence that all three strings are non-null and different from each
other), the string length does not change (i.e. there is no TICK). 
Interpreted as a three-leg Feynman diagram (a story
we cannot develop to any great extent in this paper), ADJOIN can be shown
to correspond to a ``vacuum fluctuation'' which conserves
(relativistic) 3-momentum but not energy, and hence is unobservable as
a physical process. On the other hand, when two indistinguishable
strings are compared, producing a TICK, this can be interpreted as 
four-leg Feynman diagram in which one of the two indistinguishable
strings was produced earlier and the other serves as the needed 
spectator in any observable {\it relativistic finite particle number}
three body scattering process \cite{Noyes&Jonessub}. 

Program Universe 2 also provides
a separation into a conserved set of ``labels'', and a growing set of
``contents'' which can be thought of as the space-time ``addresses''
to which these labels refer. To see this, think of all the left-hand,
finite length $S$ portions of the strings which exist when the program
TICKs and the string-length goes from $S$ to $S+1$. Call these {\it
labels} of length $L=S$, and the number of them at the critical 
tick $N_0(L)$. Further PICKs and
TICKs can only add to this set of labels those which can be produced
from it by pairwise discrimination, with no impact from the (growing in length
and number) set of content labels with length  $S_C=S-L > 0$. 
If $N_I \leq N_0(S_L)$ of these labels are {\it discriminately
independent}, then the maximum number of distinct labels they
can generate, no matter how long program universe runs,
will be $2^{N_I} - 1$, because this is the maximum number of ways we can
choose combinations of $N_I$ distinct things taking them $1,2,...,N_I$  
times. We will interpret this fixed number of possibilities as a 
representation of the  quantum numbers of systems of
 ``elementary particles'' allowed in our bit-string universe 
and use the growing content-strings to represent 
their (finite and discrete) locations
in an expanding space-time description of the universe.

This label-content schema then allows us to interpret the events
which lead to  TICK as four-leg Feynman diagrams representing
a stationary state scattering process. Note that for us to
find out that the two strings found by PICK are the same,
we must either pick the same string twice or at some previous step 
have produced (by discrimination) and adjoined the string
which is now the same as the second one picked. Although
it is not discussed in bit-string language, a little thought
about the solution of a relativistic three body scattering
problem Ed Jones and I have found \cite{Noyes&Jonessub}
shows that the driving term ($>-<\atop -$) is always a four-leg Feynman
diagram ($>-<$) plus a spectator ($\ - \ $) whose quantum numbers are 
{\it identical} with the
quantum numbers of the particle in the intermediate state 
connecting the two vertices. The 
step we do not take here is to show that the labels do
indeed represent quantum number conservation and the contents
a finite and discrete version of relativistic 
energy-momentum conservation. But we hope that enough has been said
to show how we could interpret program universe as representing
a sequence of contemporaneous scattering processes, and an algorithm
which tells us how the space in which they occur expands..  

\subsection{Cosmological Interpretation of Program Universe}

At this point we need a guiding principle to show us how we can
``chunk'' the growing information content provided by discriminate
closure in such a way as to generate a hierarchical representation
of the quantum numbers that the label-content schema provides.
Following a suggestion of David McGoveran's \cite{McGoveran98},
we note that {\it we can guarantee that the representation has a
coordinate basis and supports linear operators
by mapping it to square matrices}. 

The mapping scheme originally used
by Amson, Bastin, Kilmister and Parker-Rhodes \cite{Bastin66}
satisfies this requirement. This scheme requires us to
introduce the multiplication operation ($0\cdot 0=0=0\cdot 1=0=1\cdot
0$, $1\cdot 1=1$), converting our bit-string formalism into the {\it
field} $Z_2$. First note, as mentioned above, that any set of $n$
discriminately independent ({\it d.i.}) strings will generate exactly $2^n -1$
discriminately closed subsets ({\it dcss}). Start with two d.i. strings ${\bf a}$,
${\bf b}$. These generate three d.i. subsets, namely        
$\{ {\bf a} \}$, $\{ {\bf b} \}$, 
$\{ {\bf a},{\bf b},{\bf a}\oplus {\bf b} \}$. Require each dcss 
(\{ $\ $ \}) to contain only the eigenvector(s),
of three $2\times 2$ {\it mapping matrices} which 
(1) are non-singular (do not map onto zero) and (2) are d.i.
Rearrange these as strings. They will then generate seven dcss.
Map these by seven d.i. $4\times 4$ matrices, which meet the same
criteria (1) and (2) just given. Rearrange these as  seven d.i. strings
of length 16. These generates $127=2^7-1$ dcss. These can be mapped
by 127 $16\times 16$ d.i. mapping matrices, which, rearranged
as strings of length 256, generate
$2^{127}-1 \approx 1.7\times 10^{38}$ dcss. But these cannot be
mapped by $256\times 256$ d.i. matrices because there are at most
$256^2$ such matrices and $256^2 \ll 2^{127}-1$. 
Thus this {\it combinatorial hierarchy} terminates at the fourth
level. The mapping matrices are not unique, but exist, as has been
proved by direct construction and an abstract proof \cite{Bastinetal79}.
It is easy to see that the four level hierarchy constructed by these
rules is {\it unique} because starting with d.i. strings of length 3 or
4 generates only two levels and the dcss generated by d.i. strings of
length 5 or greater cannot be mapped. 

Making physical sense out of these numbers is a long story \cite{Noyes97a},
and making the case that they give us the quantum numbers of the
standard model of quarks and leptons with exactly 3 generations
has only been sketched \cite{Noyes94}. However we do not require
the completely worked out scheme to make interesting cosmological
predictions.  The ratio of dark to ``visible''  (i.e. electromagnetically
interacting) matter is the easiest to see. The electromagnetic
interaction first comes in when we have constructed the first three
levels giving 3+7+127 =137 dcss, one of which is identified with
electromagnetic interactions because it occurs with probability 
$1/137 \approx e^2/\hbar c$. But the construction must first complete
the first two levels giving 3+7=10 dcss. Since the construction is
``random'' and this will happen
many, many times as program universe grinds along, we will
get the 10 non-electromagnetically interacting labels 127/10 times
as often as we get the electromagnetically interacting labels.
Our prediction of $M_{Dark}/M_{B}= 12.7$ is that naive. We discuss
how we might improve the calculation of this number in the concluding section.   
 
The $1/256^4$ prediction for $N_{B}/N_{\gamma}$ is comparably naive.
Our partially worked out scheme of relating bit-string events
to particle physics \cite{Noyes94,Noyes97a}, makes it clear that
photons, both as labels (which communicate with particle-antiparticle
pairs) and as content strings will contain equal numbers of zeros
and ones in appropriately specified portions of the strings.
Consequently they can be readily identified as the most probable 
entities in any assemblage of strings generated by flipbit.
This scheme also makes the simplest representation of fermions and
anti-fermions contain one more ``1'' or one more ``0'' than the photons.
(Which we call ``fermions'' and which ``anti-fermions'' is, to begin
with, an arbitrary choice of nomenclature.) Since our dynamics insures
conventional quantum number conservation by construction, the problem
--- as in conventional theories---is to show how program universe
introduces a bias between ``0'' 's and ``1'' 's
once the full interaction scheme is developed. (The recently commissioned 
``B-Factory at SLAC is aimed at providing experimental evidence relevant 
to a conventional  explanation of the observed bias between matter and anti-matter in
our universe.) 

\begin{figure}[htb]
\begin{center}
\leavevmode
\epsfbox{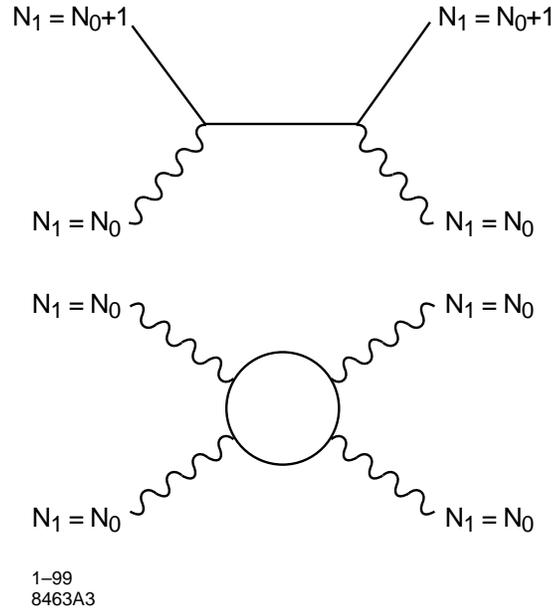}
\end{center}
\caption[*]{Comparison of bit-string labeled processes after the label
length is fixed at 256 interpreted as baryon ($N_1=N_0+1$) 
photon ($N_1=N_0$) and photon-photon scattering. Here $N_1$ 
and $N_0$ symbolize, respectively,  the number of ones and zeros in the 
label part of the string  (which is of length 256). Program universe guarantees 
that, in the absence of further considerations, the content part of the strings 
will have an equal number of zeros and ones with very high probability as 
the string length (universe) grows. }
\label{fig1}
\end{figure}

Since program universe has to start out with two
strings, and both of these cannot be null if the evolution is lead
anywhere, the first significant PICK and discrimination will
necessarily lead to a universe with three strings, two of which are
``1'' and one of which is ``0''. Subsequent PICKs and TICKs are
sufficiently ``random'' to insure that (at least statistically)
there will be an equal number of zeros and ones, apart from the initial
bias giving an extra one. Once the label length of 256 is reached,
and sufficient space-time structure (``content strings'') generated and
interacted to achieve thermal equilibrium, this label bias for a 1
compared to equal numbers of zeros and ones will persist for 1 in
256 labels. But to count the equilibrium processes relevant 
to computing the ratio of baryons to photons, we must compare the
labels leading to baryon-photon scattering compared to those
leading to photon-photon scattering. This requires the
baryon bias of 1 to appear in one and only one of the four labels
of length 256 involved in that comparison; this comparison
is illustrated in Fig. \ref{fig1}, which assumes that the above mentioned
interpretation of the strings causing observable TICK's as
four  leg Feynman diagrams has been satisfactorily demonstrated. 
We conclude that, in the absence of further information, 
$1/256^4$ is the program universe prediction
for the baryon-photon ratio at the time of big bang nucleosynthesis.
We will discuss how this estimate could be strengthened and refined in
the concluding section.   

\subsection{Comparison with Observation}

The currently accepted way to set the stage for the discussions of 
cosmology is to note that if the universe is homogeneous and
isotropic on a large enough scale (for which hypothesis there is now 
claimed to be good evidence) and postulate Einstein gravitation 
(at least
in the weak field limit, for which there is again good evidence)
the Friedman-Robertson-Walker (FRW) equations apply. Further,
if we know the boundary conditions at the time of big-bang
nucleosynthesis (when the event horizon was ``only'' a million or so
times smaller than it is now), we can integrate these equations up
to the current time knowing only the two parameters $\Omega_M$ and
$\Omega_{\Lambda}$ \cite{Olive98}. Here $\Omega_M =\rho_M/\rho_c$
is the ratio of the contemporary density of matter to the critical
density in the absence of a cosmological constant
($\Omega_{\Lambda}=0$). We can get this knowing only Newton's
gravitational constant $G_N$ and the current value of the Hubble
constant $H_0\equiv 100 \ h_0 \ km \ s^-1 Mpc^-1$ since
$\rho_c= 3H_0^2/ 8\pi G_N = 1.88 \times 10^{-29} h_0^2\ g \ cm^{-3}$.
Similarly the scaled cosmological constant is given by
$\Omega_{\Lambda} = \Lambda c^2/3H_0^2$, where $\Lambda$ is the
integration constant which must appear in solving the FRW differential
equations.

\begin{figure}[htb]
\begin{center}
\leavevmode
\epsfbox{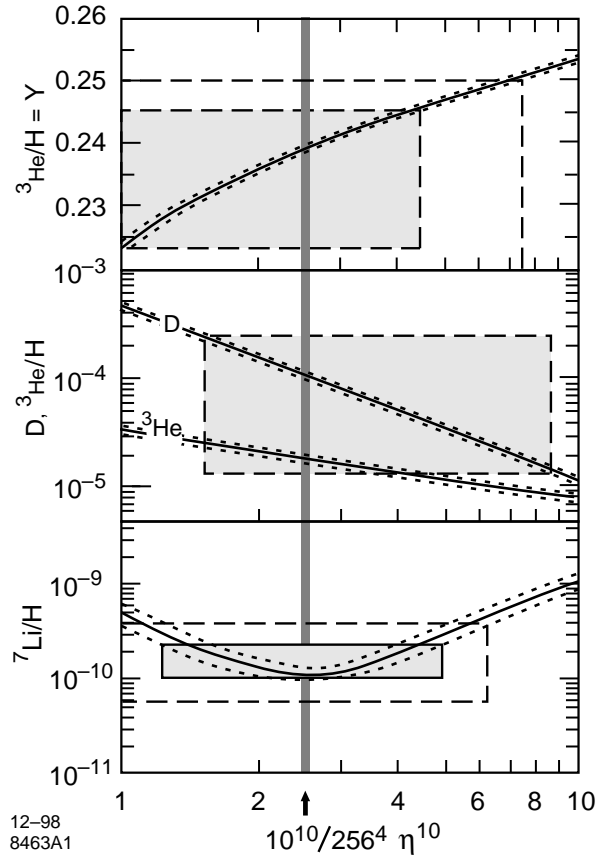}
\end{center}
\caption[*]{Comparison of the bit-string physics prediction that $\eta=
256^{-4}$ with accepted limits on the cosmic abundances as given by
Olive and Schramm in  \cite{PDG98}, p.~119.}
\label{fig2}
\end{figure}

\begin{figure}[htbp]
\begin{center}
\leavevmode
\epsfbox{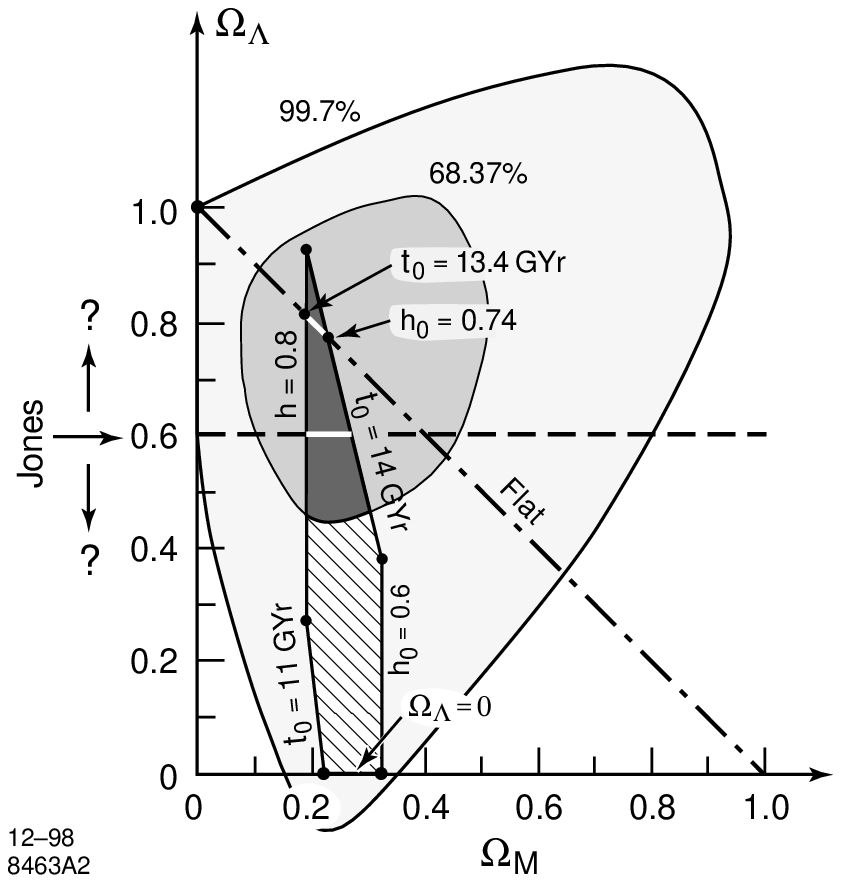}
\end{center}
\caption[*]{Limits on $(\Omega_M, \Omega_{\Lambda})$  set by combining
the Supernovae Type Ia data from Perlmutter, et al. with the Cosmic
Ray Background Experiment (COBE) satellite data as quoted by Glanz
 \cite{Glanz98a} (dotted curves at the 68.37\% and 99.7\% confidence
levels) with the predictions of bit-string physics that $\eta_{10} = 10^{10}/256^4$ 
(cf. Fig. \ref{fig1}) and $\Omega_{\rm Dark}/\Omega_B =12.7$. We accept the constraints on 
the scaled Hubble constant $h_0=0.7\pm 0.1$  \cite{Groometal98} and 
on the age of the universe $t_0 = 12.5\pm1.5 \ Gyr$ (solid lines).
We include the predicted constraint $\Omega_{\Lambda} > 0)$. The Jones estimate
of $\Omega_{\Lambda}= 0.6$ is indicated, but the uncertainty is not available. }
\label{fig3}
\end{figure}

The two {\it program universe} results we consider here are that
the ratio of dark to baryonic matter is 12.7 to 1 and that the
ratio of baryons to photons at the time of nucleosynthesis,
symbolized by $10^{-10}\eta_{10}$, is $1/256^4$. Our naive arguments
for these numbers are given in the last section; from now on we accept
them as predictions to be tested. We show in Fig. \ref{fig2} that this
value for $\eta_{10}$ well represents a central value consistent
with the cosmic abundances of the light elements. 
Since we know 
from the currently observed photon density (calculated from the
observed $2.782 \ ^oK$ cosmic background radiation) 
that the normalized baryon density is given by \cite{Olive98}
\begin{equation}
\Omega_B = 3.67\times 10^{-3}\eta_{10}h_0^{-2}
\end{equation}
and hence from our assumption about dark matter that the total mass 
mass density will be 13.7 times as large, we have that
\begin{equation}
\Omega_M= 0.11706 h_0^{-2} \ .
\end{equation}   
Hence, for $0.8 \geq h_0 \geq 0.6$  \cite{Hogan98}, $\Omega_M$ runs from
$0.18291$ to $0.32517$. This clearly puts no restriction
on $\Omega_{\Lambda}$. 

Our second constraint comes from integrating the scaled
Friedman-Robertson-Walker (FRW) equations from a time 
after the expansion becomes matter dominated with no pressure
to the present. Here we assume that this initial time is 
close enough to zero on the time scale of the integration
so that the lower limit of integration can be approximated 
by zero \cite{Primackinp}.
Then the age of the universe as a function of the current values of
$\Omega_M$ and $\Omega_{\Lambda}$ is given by
\begin{eqnarray}
t_0&=&9.77813 h_0^{-1}f(\Omega_M,\Omega_{\Lambda}) \ Gyr\nonumber\\
&=&9.77813 h_0^{-1}f(0.11706h_0^{-2},\Omega_{\Lambda}) \ Gyr
\end{eqnarray} 
where
\begin{equation} 
f(\Omega_M,\Omega_{\Lambda})=\int_0^1dx 
\sqrt{{x\over \Omega_M +(1-\Omega_M-\Omega_{\Lambda})x 
+\Omega_{\Lambda}x^3}} \ .
\end{equation}
For the two limiting values of $h_0$, we see that
\begin{eqnarray}
h_0&=&0.8,\ \ t_0=12.223f(0.18291,\Omega_{\Lambda}) \ \ Gyr \nonumber\\  
h_0&=&0.6,\ \ t_0=16.297f(0.32517,\Omega_{\Lambda}) \ \ Gyr \ .
\end{eqnarray}
The results are plotted in Fig. \ref{fig3}.

To orient ourselves in the $(\Omega_M,\Omega_{\Lambda})$ plane, we
first consider a flat universe in which the curvature term in the
normalized FRW equations vanishes, i.e.
$1-\Omega_M-\Omega_{\Lambda}=0$. Then for $h_0=0.8$, performing the
integration, these constraints predict $t_0=13.8 \ Gyr$, barely within the
allowed range. The upper limit of $14 \ Gyr$ requires that $h_0= .737, \Omega_M =
.199, \Omega_{\Lambda} = .801$. We conclude that requiring flatness 
together with $\eta_{10} = 2.33$ and  $M_{\rm Dark}/M_B=12.7$
restricts us to the short line segment from
$(\Omega_M,\Omega_{\Lambda})= (0.183,0.817)$ to $(0.199,0.801)$.
At the same time, this sets a lower bound of $0.737$ on $h_0$,
and of $13.8 \ Gyr$ on $t_0$.
It is therefore very important for our bit-string cosmology to 
know how flat it has to be. Otherwise we may soon be forced to modify or abandon 
this approach to cosmology as the observational data improve.  

If we relax the flatness assumption, but take from our model (see
Section 3.1) the requirement that the cosmological constant be repulsive
(space generates more space as time goes by), 
the predicted limits on our parameters
as plotted in Fig. \ref{fig3} are well within the 99.7\% confidence
limit given by putting together
the type Ia supernovae and the COBE data \cite{Glanz98a,Glanz98b}.
At the 68.37\% confidence limit provided by this data
we see that, if we can find a way to justify our choice for $\eta_{10}$
and the ratio of dark to baryonic matter, we require the
cosmological constant to lie between 0.45 and 0.94.
We conclude that the bit-string cosmology is within the observational
bounds, and that either a calculation of the limits on flatness
or of the limits on the cosmological constant would greatly improve
the predictive power of our theory. 

\section{Jones' Cosmological Constant}

Since Jones' paper \cite{Jones98} is still not submitted, I am at liberty here only to
quote the following sentence
\begin{quote}
From general operational arguments, Ed Jones has shown how to start
from $\sim N$ Plancktons and self-generate a universe with $\sim N'$
baryons which---for appropriate choice of $N$---resembles our
currently observed universe. In particular it must necessarily have a 
positive cosmological constant characterized by $\Omega_{\Lambda} \sim
0.6$ 
\end{quote} 

We note further that Jones' general arguments a) are completely
compatible with {\it program universe} and b) do not in themselves
fix the value of $N$. Further, the estimate given above, which
was made
before and independent of the calculations reported in the last
section, falls squarely in the middle of the allowed region
(see Fig. \ref{fig3}). Clearly, pursuing the combination of these two lines of reasoning 
could prove to be very exciting. We indicate how this might be done in
the concluding section. 

\section{A Research Proposal for ANPA}

We believe that the above calculations amply justify our contention
made in the introduction
that {\it if} the ANPA program can be shown in a convincing way to 
lead to the 
prediction of the two parameters $\eta_{10}$ and $\Omega_{\rm Dark}/\Omega_B$
to anything like the precision that $\hbar c/e^2$ and
$[M_{\rm Planck}/m_{\rm proton}]^2$ are given in the lowest approximation
by the older triumph \cite{Bastin66} {\it then}
we can provide a target for the observational cosmologists to shoot at.
If that happens, as observations improve, we will be either vindicated or
shown to have made some fatal flaw in our reasoning. This is a much
more exciting game to play than trying to show physicists that we 
can approximately compute numbers that they already are confident
they can measure to higher precision than we can provide. 

The problem is that the naive arguments given above for the numbers
studied here are---even within ANPA---unlikely to be convincing
to any one other than a sympathetic reader. A friendly critic would
at best characterize them as ``heuristic'' and a less friendly critic
as ``hand-waving''. A hostile critic will dismiss them as ``wishful
thinking.''  I readily admit that we need to do better, but fear that
the amount of work needed is beyond my reach. Fortunately, most of it
is precisely what needs to be done in any case, if the elementary
particle end of the ANPA program is not to stagnate. I now outline
a possible research strategy.

I propose that we first construct a rigorous bit-string theory for
renormalized QED in the truncated version of a single
particle-antiparticle mass and the first 
combinatorial hierarchy approximation for 
the fine structure constant, i.e. 1/137. Basically, I believe this only
involves putting together two pieces of the puzzle
which have already been completed, and which we now discuss. 

Following a
suggestion of Feynman's \cite{Feynman&Hibbs65}, Lou Kauffman and I
have shown \cite{Kauffman&Noyes96b} that, given as the boundary
condition a rational fraction velocity between two fixed end-points
in 1+1 dimensional space-time,  a finite and discrete version of the free particle
Dirac Equation can be solved  by an appropriate collection of
bit-strings pairs interpreted as ``random walks''. Hence, once we have 
shown how to couple the beginning- and end-points to bit strings
representing photons which satisfy the appropriate conservation laws
in three dimensions, the ``renormalized single particle propagator''
problem for fermions will have been solved \cite{Noyes95}.

Again following an idea originally due to Feynman \cite{Feynman48},
as presented by Dyson \cite{Dyson90} and developed further by Tanimura
 \cite{Tanimura92}, Kauffman and I have shown \cite{Kauffman&Noyes96a}
that the discrete physics hypothesis that first measuring position
and then velocity is different from first measuring velocity and then
position leads to the {\it relativistic} commutation relations
needed to undergird, rigorously, the Feynman-Dyson-Tanimura ``proof''
of the free particle Maxwell Equations using in addition only Newton's
second law connecting force to field. This amounts to (for a single
particle trajectory) emission or absorption of ``photons'' at
finite and discrete points in 3-space connected by straight
line segments. We have noted above that these in turn can be
represented by collections of random walks of a Dirac particle.
What remains is to show that the ``interaction'' so described
does indeed consistently describe the connection between bit-string
photons and bit-string Dirac particles in the $\hbar c/e^2 =137$
third level hierarchy calculation. That this way of looking at the
hydrogen atom also provides the {\it relativistic} connection between
binding energy, principle quantum number and coupling constant first
given by Bohr \cite{Bohr15} has already been
proved \cite{McGoveran&Noyes91}. Bits and pieces of the geometrical
interpretation of the angles between bit-strings needed to lace
all this together also exist \cite{Noyes97a}. 

I feel that a concerted
effort could get to a lowest order renormalized QED in this way,
providing the bit-string underpinning for the renormalized Feynman
diagrams needed to discuss the equilibrium between protons and
black body gamma radiation (Compton scattering and photon-photon
scattering) to a part in 137 at the time of ``big-bag
nucleosynthesis''. The black-body spectrum itself is guaranteed
by the statistical character of string creation in program universe
and the indistinguishability (in the usual sense of Bose-Einstein
statistics) of the photon states in a bit-string representation of
photons. If the bit-string version of the finite particle number
relativistic scattering theory we have started to
construct \cite{Noyes&Jonessub} and the quantum numbers of the standard
model we have sketched \cite{Noyes94} are not sufficient to describe
the nuclear physics to the level needed at the time of
big bang nucleosynthesis and later,
our cosmology obviously cannot get off the ground. This is
the reason why, up to now, I have given priority to putting the
elementary particle physics and scattering theory on a firm foundation.

The next step, as I see, it is to make a more careful analysis (or
possibly to run computer experiments) to find out reliably---%
rather than heuristically---what the probable
distribution of label and content strings generated by program universe
must be. This may actually help in getting the quantum number
interpretation of the bit-string scattering theory straight.
If this does {\it not} end up giving something close to $1/256^4$
for $\eta$, either program universe will have to be modified,
or the whole bit-string cosmology abandoned.  
 
These steps in turn are needed---but presumably at a much earlier stage
in the string evolution described by program universe than we have 
been discussing above---in order to gain confidence in the prediction
of 12.7 for the dark matter-baryon ratio.  Here I foresee two ways to 
go. One is to revive an old idea of Wheeler's \cite{Wheeler55}: {\it
Geons}. These are classical configurations of electromagnetic waves
which are so energetic that their mass is sufficient to bind them
together gravitationally as standing waves. Within classical physics,
Wheeler showed that they are indeed stationary solutions of the coupled
Einstein and Maxwell equations in the absence of particles. Because
they are classical, they can be of any size thanks to scale invariance;
the only dimensional constants which occur in the theory are $G_N$ and
$c$. But once one includes Planck's constant, breaking scale
invariance, Wheeler showed that quantum effects start to become
important already when the masses are many times the range of stellar
sizes, and rapidly become dominant at smaller scales. Thus there
were no observed candidates for such objects when the paper was
published. But now that we know that there are enormous distributions
of Dark Matter of the size of clusters of galaxies \cite{Tyson98},
we do have observational evidence that might be relevant.

The problem, as with particulate dark matter (which we discuss below),
is to see how program universe might be expected to generate such
structures. In the completed label scheme the string which interacts
with {\it everything} is the unique label of length 256 which contains
256 ones (the {\it anti-null string}). This will represent the
Newtonian static gravitational interaction. The combinatorial
hierarchy construction shows that for protons this corresponds to a 
dimensionless coupling constant $G_Nm_p^2/\hbar c \approx 2^{-127}$.
But at earlier levels in the construction the analogous anti-null string
occurs with probabilities $1/3, 1/10, 1/137$ as
levels 1, 2 and 3 are completed. Thus the equivalent of a very strong
gravitational interaction occurs during very early stages of the
construction. This will bind together electrically neutral objects,
some of which will continue to be electromagnetically neutral as the
strong, electromagnetic, and weak interactions evolve toward their
final form. These early objects might end up as something like enormous
geons as the construction proceeds. The idea looks to me to be work
exploring both in classical and in bit-string physics.

For the particulate dark matter, we also have a   
class of candidates. In our unsuccessful attempt to get our 
views into {\it Physical Review Letters} \cite{Noyes91},
we pointed out that a proton together with
$2^{127}$ gravitating proton-antiproton pairs assembled with spin 1/2 
within a radius of $\hbar/m_pc$  
would form a ``charged, rotating black hole'' with Beckenstein
number $2^{127}$. It would then rapidly decay by Hawking radiation,
but in our theory, since baryon number is conserved, would
leave behind a proton. Therefore we claim that in our theory
the proton is ``gravitationally stabilized''. Although we did
not point it out in that note, the same argument stabilizes an
electron, due to charge and lepton number conservation, and also stabilizes 
a (massive) electron-type neutrino, due to lepton number conservation. 
The neutral current interaction will, of course, provide the
neutrino with a finite mass once the label-content assemblage
has developed far enough. So our theory will provide neutral
assemblages of photons, gravitons and (electron-type) neutrinos
which bid together gravitationally. These will be our candidates
for particulate dark matter.

In both cases we need to (a) work out the actual models for this
neutral dark matter and (b) show that the 127/10 argument does apply to
them when we have studied program universe in more detail. 
For the particulate types, we will also be under the obligation
to calculate detection cross sections and show that extant
dark matter searches would not have picked them up. Of course,
if we are very lucky, we might be able to suggest new types of
searches that would pick up our candidates, if they are there.

To complete the task we must, minimally, show in detail how the Jones
argument (cf. Section 4) applies to program universe. This does not appear
to be too difficult. Better, by examining program universe in more
detail we might provide a statistical law as how the ratio between
space and matter evolves. This could then form the basis for an actual
calculation of the cosmological constant in the most probable of all
universes. Whether this is also the best of all universes we will leave
to the theologians to argue.

I hope I have said enough to indicate some of the exciting things that 
attention to cosmology could open up for the ANPA program in the
future. I can only hope that I will be around long enough to see  
some of them bear fruit.

In closing I wish to thank Ed Jones for several illuminating
discussions of cosmology before and after ANPA 20, and Brian Koberlein
for checking out my understanding of the implications of current
observations at ANPA 20 before I made my presentation. Of
course I am responsible for any errors that may have crept in.

\end{document}